\documentclass[a4paper]{article}

\usepackage{INTERSPEECH2022}
\usepackage{xcolor,listings}
\usepackage{amsmath,graphicx,amssymb}
\usepackage{hyperref}
\hypersetup{
    pdftitle={Extending RNN-T based speech recognition systems with emotion and language classification},
    }
\usepackage{cite}
\usepackage{booktabs,url}
\usepackage{booktabs}
\usepackage{multirow}
\usepackage{color,soul}

\definecolor{backcolour}{rgb}{0.95,0.95,0.92}

\lstdefinestyle{mystyle}{
    backgroundcolor=\color{backcolour}
}

\title{Extending RNN-T-based speech recognition systems with emotion and language classification}
\name{Zvi Kons, Hagai Aronowitz, Edmilson Morais, Matheus Damasceno, Hong-Kwang Kuo, Samuel Thomas, George Saon}
\address{IBM Research AI}
\email{\textcolor{white}{hidden.}zvi@il.ibm.com\textcolor{white}{.hidden}}

\begin{document}

\maketitle
\begin{abstract}
Speech transcription, emotion recognition, and language identification are usually considered to be three different tasks. Each one requires a different model with a different architecture and training process. We propose using a recurrent neural network transducer (RNN-T)-based speech-to-text (STT) system as a common component that can be used for emotion recognition and language identification as well as for speech recognition. Our work extends the STT system for emotion classification through minimal changes, and shows successful results on the IEMOCAP and MELD datasets. In addition, we demonstrate that by adding a lightweight component to the RNN-T module, it can also be used for language identification. In our evaluations, this new classifier demonstrates state-of-the-art accuracy for the NIST-LRE-07 dataset.
\end{abstract}

\section{Introduction}
Human speech is a rich signal that carries with it a vast amount of information. In addition to words, the speech signal encodes information about the speaker's identity, language, accent, emotions, and physical state. The human brain is tuned to extract and combine all these types of information from the speech signal. Yet, from a machine learning perspective, recognizing each of these parts is considered an independent challenging problem. Usually, a separate model is trained to extract information from each part of the speech data.

With the recent advances in self-supervised learning for speech, there are many general purpose models available which have been pre-trained on large speech corpora. Such models are more easily used for several different tasks \cite{SUPERB,csae}. Nevertheless, in most cases, the best results are achieved by fine-tuning the model to a specific task.

Automatic speech recognition (ASR) models based on RNN-T architecture are also becoming common \cite{rnnt-graves}. These models are  trained on large speech corpora, but in a supervised, task-oriented manner. The main task of these models is speech transcription, but there have been some recent attempts to expand them to incorporate other tasks such as diarization \cite{Diarization} and intent classification \cite{convy2}.

In this paper, we present a practical guide to extending an ASR-based RNN-T model to perform two additional tasks: speech emotion recognition (SER) and language identification (LID). We explain how we added these two features without affecting the performance of the original speech transcription functionality. 

An ASR system that can perform several tasks at once, such as the one we propose,  has many advantages over several single-task systems. First, it can generate a rich transcript. This can be helpful in many scenarios, such as call center interactions, where understanding users' emotions and identifying their language correctly can improve the interaction. From a deployment point of view, having a single model that can perform multiple tasks is a huge advantage since it requires fewer resources and less maintenance.

State-of-the-art audio-based SER classifiers use audio encoding models to first extract frame-level features and then apply a classifier on them \cite{PORIA201798}. These audio-based classifiers can benefit from additional information that can be extracted from semantic text analysis models \cite{ed-ser}.

We added the SER functionality to an STT model by training it to output additional emotion tags along with the text transcription; this is somewhat similar to what was done in \cite{convy2}. Using an RNN-T model for SER is very advantageous because it has both an audio analysis network and a text prediction network. This allows it to leverage multi-modal information from both audio and text when it generates the emotion tags. The SER models and the related experiments are described in Section \ref{sec:ser}.

The second part of our work focuses on automatic speech recognition for audio samples, where the underlying language is not known a priori. The commonly used approach to this problem is to first use an LID classifier to identify the correct language and then apply the corresponding, single-language STT model. Our goal was to simplify this process into a single step.

Another option for speech recognition is to use multilingual STT models. These models \cite{microsoft-multi,amz-BilingualASR,google-ML} can be  trained on more than one language so they are able to identify the language and transcribe it at the same time. Multilingual STT systems can be harder to train than single language models and can require additional resources. In many cases, these models depend on internal or external LID classifiers.
In addition, multilingual STT models require transcribed speech samples for training in all of their languages. This can lead to problems when a language has very few resources of this kind to use for training. 

In Section \ref{sec:LID}, we describe a new method for LID classification that is based only on the audio encoder part of a single-language RNN-T model. Using this model, we show state-of-the-art results for the NIST-LRE-07 dataset. With this approach, it is much easier to train the LID classifier because we do not need transcribed audio. We can also identify the language quickly without waiting for the text decoding and do this with only minimal overhead. This allows the addition of LID functionality to existing ASR systems, without any change to their STT models.

\section{Joint ASR and speech emotion recognition}
\label{sec:ser}
Our first step was to extend an RNN-T model \cite{ibm-stt} to produce both a transcript of the speech and emotion tags. We started with a pre-trained US-English model that is trained to output the transcript as a sequence of symbols representing alphabetical characters. We extended the output symbols by adding a new symbol for each one of the emotions we wanted to identify. In the transcript, these symbols are translated into unique emotion tags. For example, an output transcript might look like this: 

\begin{lstlisting}
I FEEL HAPPY TODAY <HAPPY>
\end{lstlisting}

\noindent where  ``$<$HAPPY$>$'' is the emotion tag produced by the symbol that indicates happy speech.

In the training stage, for each utterance, we append the corresponding emotion tag to the end of its target text. The location of the tag within the text is important. While the RNN-T encoder is bidirectional, the prediction network is unidirectional. By placing the tag at the end of the utterance, we allow the system to accumulate information from both the speech and the transcribed text before it makes a decision. We trained the system to output both the target text and the tag at the end. We did this using the RNN-T loss function in the same way as it was initially trained to output only the transcript \cite{ibm-stt}.

In the course of our work, we found that the values of the loss function were not good indicators for the accuracy of the emotion classification. Consequently, we calculated the accuracy of the emotion classification over a development set, and used this as a stop criterion and for best model selection.

During inference, we used our modified STT model to generate the text transcript with the emotion tags. For each utterance, we searched the output text for the last emotion tag that was output. This tag was then used as the predicted emotion for the utterance. If no tag was output, we labeled the utterance as neutral.

\subsection{Experimental evaluation}
\subsubsection{IEMOCAP dataset}
Our first experiment was done with the IEMOCAP dataset \cite{IEMOCAP}. The dataset is split into five sessions, which allowed us to do a five-fold cross validation by training on four sessions and testing on the fifth. Out of the seven emotions labeled in the dataset, we used only four: neutral, happy, angry, and sad.
We randomly split the training data into training and development sets. The development set was used during the training for the stop criterion and for selecting the best model.

We measured the emotion classification accuracy over the test set and calculated the character error rate (CER) and word error rate (WER) from the generated transcript. We then compared the CER and WER to the baseline model to verify that our training did not make the transcription less accurate.

The results of the experiments are presented in Table \ref{tab:iemocap}. The first line shows the baseline CER and WER for the speech transcription, using the baseline STT before any of our training. The second line shows the results after we fine tuned the whole model. As can be seen, the new model provides good accuracy on emotion classification while also improving the transcription accuracy.

In the next experiment, we repeated the training but the acoustic encoding network was frozen, so only the prediction and joint networks were tuned. This is somewhat similar to what we describe in Section \ref{sec:LID} where we use the encoder for LID. This experiment was designed to examine whether we could attain good emotion classification without acoustic adaptation to the speech in the dataset. The results show a considerable drop of more than 10\% in accuracy.

For comparison, we also include state-of-the-art results that were achieved on the same data, using LID classifiers build on top of the wav2vec 2.0 and HuBERT encoders, reported in \cite{ed-ser} and \cite{csae}. These results were achieved by a single task systems that were trained to produce only emotion classification. As can be seen, the results we report in this paper are close, but there is still a gap to the state-of-the-art models.

The IEMOCAP data is a combination of scripted and improvised sessions. In the scripted part, many phrases are common to different sessions. We wanted to verify that the classifier does not simply learn the emotions from the text, so we repeated the above experiments on each type of session independently. The results are also presented in Table \ref{tab:iemocap}. We found only a small difference in the accuracy of the emotion classification between the two session types. This means that the models learn relevant audio features and not just the text.

\begin{table}[htb]
\centering
\resizebox{\linewidth}{!}{%
\begin{tabular}{@{}lllll@{}}
\toprule
\textbf{Dataset}                     & \textbf{Experiment}     & \textbf{Emotion (\%)} & \textbf{CER (\%)} & \textbf{WER (\%)} \\ \cmidrule(l){2-5} 
\multirow{3}{*}{\textbf{Full}}       & \textbf{Baseline STT}   &                           & 17.0       & 24.2       \\
                                     & \textbf{Whole model}    & 72.0                    & 11.4       & 20.8       \\
                                     & \textbf{Encoder frozen} & 58.2                    & 15.7       & 26.7       \\
                                                                          & \textbf{wav2vec 2.0} & 76.5                    &        &        \\
                                                                          & \textbf{HuBERT} & 75.2                    &        &        \\
                                         \cmidrule(l){2-5} 
\multirow{2}{*}{\textbf{Scripted}}   & \textbf{Baseline STT}   &                           & 12.5       & 19.9       \\
                                     & \textbf{Whole model}    & 71.5                    & 6.8        & 15.4       \\ \cmidrule(l){2-5} 
\multirow{2}{*}{\textbf{Improvised}} & \textbf{Baseline STT}   &                           & 20.9       & 28.1       \\
                                     & \textbf{Whole model}    & 73.1                    & 17.1       & 27.1       \\ \bottomrule
\end{tabular}
}
\caption{Experiment results for the IEMOCAP dataset}
\label{tab:iemocap}
\end{table}

\subsubsection{MELD dataset}
Our second experiment was done with the MELD dataset \cite{MELD}. Out of the seven emotions labeled in the dataset, we used only four: neutral, joy (happy), angry, and sad.
We divided the MELD dataset into train, development, and test sets following the standard data split suggested in \cite{MELD}. We repeated the same experiments measuring emotion classification accuracy, CER, and WER.

The results of the experiment are shown in Table \ref{tab:meld}. The first three lines show the same experiment as in the previous table. Again, the results show good emotion classification accuracy with an improvement in the transcription accuracy. However, there is a considerable drop of more than 4\% in the accuracy when the acoustic encoder network is not trained.

For comparison, we also include state-of-the-art results that were achieved in our previous experiments using the wav2vec 2.0 and HuBERT encoders following the same experimental method described in \cite{ed-ser}. Here too, our current accuracy is still a bit lower than the state-of-the-art.

\begin{table}[htb]
\centering
\resizebox{\linewidth}{!}{%
\begin{tabular}{@{}llll@{}}
\toprule
 \textbf{Experiment}     & \textbf{Emotion (\%)} & \textbf{CER (\%)} & \textbf{WER (\%)} \\ \cmidrule(l){2-4} 
 \textbf{Baseline STT}   &                           & 56.6       & 68.5       \\
\textbf{Whole model}    & 56.6                    & 41.1       & 41.7       \\
\textbf{Encoder frozen} & 52.1                    & 49.6       & 49.0       \\
\textbf{wav2vec 2.0} & 63.76                    &        &        \\
\textbf{HuBERT} & 62.94                    &        &        \\            
\bottomrule
\end{tabular}
}
\caption{Experimental results for the MELD dataset}
\label{tab:meld}
\end{table}

\section{Language identification}
\label{sec:LID}
\begin{figure}[htb]
  \centering
  \includegraphics[width=\linewidth]{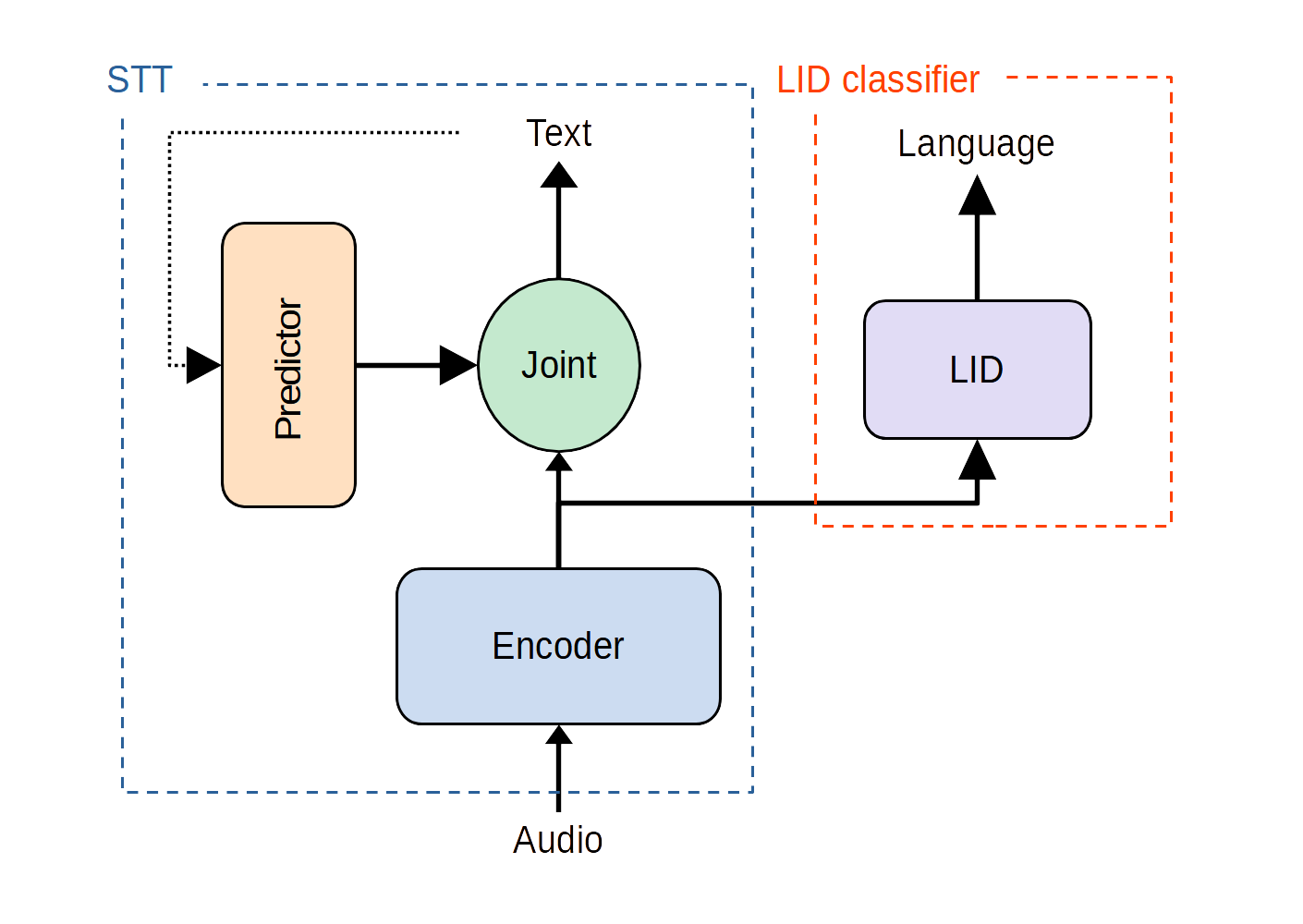}
    \caption{Structure of the LID classifier network.}
  \label{fig:rnnt-w-lid}
\end{figure}

We create a language classifier by incorporating a lightweight classifier into an RNN-T network. The classifier's input is the output from the acoustic encoder part of the RNN-T model  (Figure \ref{fig:rnnt-w-lid}). The role of the encoder network is to convert audio into frame-level embedding features that are useful for STT. This is somewhat similar to what is done by self-supervised speech representation models such as HuBERT \cite{hubert} or wav2vec 2.0 \cite{wav2vec2}. In contrast, the RNN-T encoder is trained in a supervised manner so its output features are likely to contain more phonetic and linguistic information and less information about the speaker and the acoustic channel. These features are highly useful for LID.

There are several advantages to our configuration over other types of LID classifiers or multilingual STT. First, the training process can be much faster because we already have a trained RNN-T model, so we only need to train a lightweight classifier network. Second, as we show in Section \ref{sec:lid-expr}, we do not need a multilingual STT. We can use an encoder from a single-language RNN-T model and only train the classifier on the required languages. For the classifier training, we need training samples labeled by their languages. We do not need transcriptions of those samples. This makes it much easier to obtain the required data.

Another advantage is that we do not need to run the text decoding part of the RNN-T model, which is much slower than running just the encoder. This is useful, for example, if our STT is expecting a default language. We apply the encoder on the input speech and before continuing to the decoding, we can verify that the input is in the correct language. If it is, then we can continue with the decoding using the encoder output that we already have. If not, we can then switch to an STT model for the correct language.

\begin{figure}[htb]
  \centering
  \includegraphics[width=0.7\linewidth]{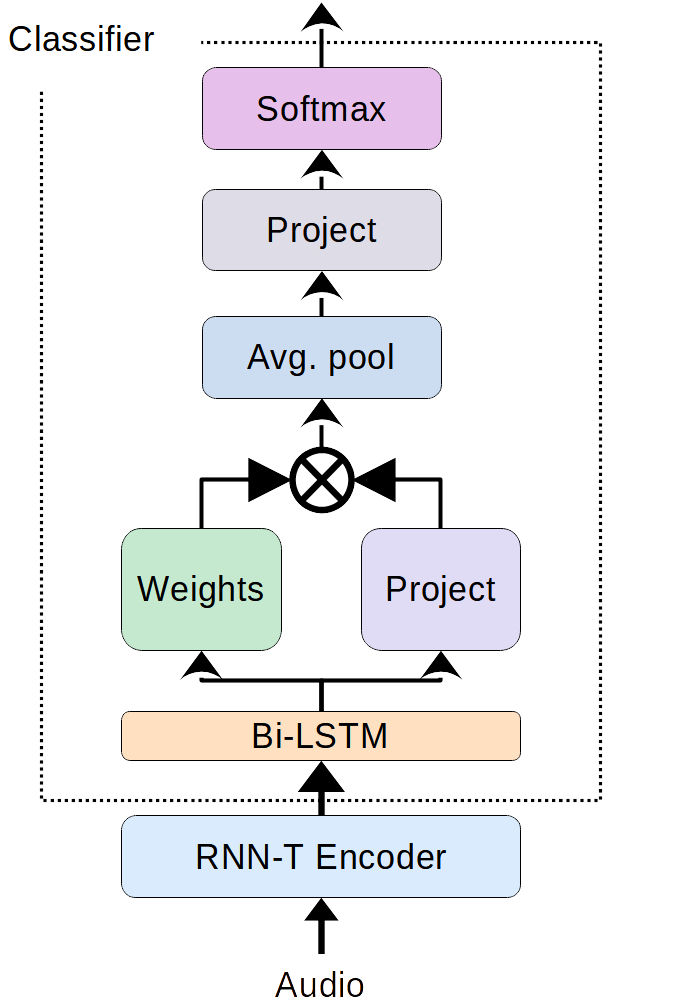}
    \caption{The structure of the classifier network.}
  \label{fig:classifier}
\end{figure}

The classifier structure is shown in Figure \ref{fig:classifier}. The output of the RNN-T encoder is processed by a bidirectional long short-term memory (LSTM) layer. This is followed by a multi-head weighted-average pooling layer:

\begin{equation}
\label{eq:weights}
w_t = \mathrm{Pr_2} \left( \mathrm{\sigma} (\mathrm{Pr_1}(x_t)) \right)
\end{equation}
\begin{equation}
\label{eq:pooling}
y = \frac {\sum_t e^{w_t} \mathrm{ReLU} (\mathrm{Pr_3} (x_t))} {\sum_t e^{w_t}}
\end{equation}

\noindent where $x_t$ is the LSTM output vector at time $t$, $w_t$ is the weight vector for this frame, $\mathrm{Pr_n}()$ are linear projections, $\mathrm{\sigma}()$ is the log-sigmoid function, and $y$ is the pooled weighted average. The final score for each language is calculated from $y$ using a linear projection and a SoftMax function.

\subsection{Experiments with the NIST-LRE-07 dataset}
\label{sec:lid-expr}
We tested our LID classifier on the NIST-LRE-07 dataset \cite{nist-lre-07}. We followed a similar procedure to the one in \cite{csae}, except we used the 8KHz audio. The dataset contains training and test data for 14 languages.

We applied the acoustic encoder part of a pre-trained, US-English, RNN-T model on each one of these samples and used the output features to train and test the classifier. Part of the training data was held out to form a validation set that was used for a stopping criterion and for selecting the best model.

Table \ref{tab:lid} reports our results measured in equal-error-rate (EER). For comparison we add two more experiments. In the first, we use the same system but this time allow fine-tuning of the RNN-T encoder network. This makes the LID classifier more accurate but now, the encoder part cannot be reused for text decoding.

\begin{figure}[htb]
  \centering
  \includegraphics[width=\linewidth]{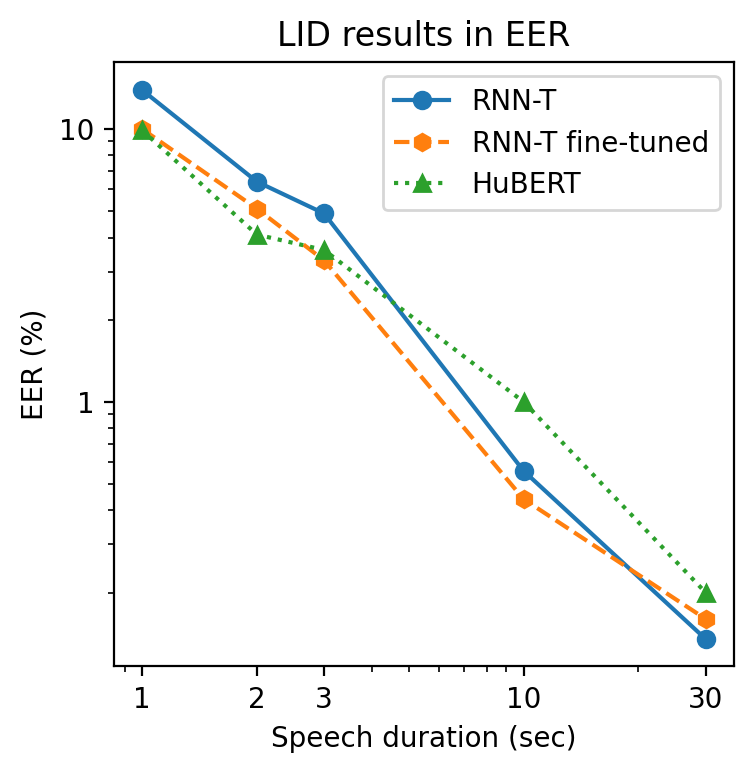}
    \caption{EER for the LID classifier vs. speech duration (log-log scale)}
  \label{fig:lid-eer}
\end{figure}

For the second experiment, we replace the RNN-T encoder network with an HuBERT network and fine-tune the whole network. This is similar to the experiments in \cite{csae} but this time we use LSTM with weighted average pooling as in \eqref{eq:weights} and \eqref{eq:pooling} instead of a simple mean pooling. For easier comparison, the same results are also shown in Figure \ref{fig:lid-eer} on a log-log scale.

\begin{table}[htb]
\centering
\begin{tabular}{@{}llllll@{}}
\toprule
\textbf{System} & \textbf{1s} & \textbf{2s} & \textbf{3s} & \textbf{10s} & \textbf{30s} \\ \midrule
RNN-T LID       & 13.9         & 6.4            & 4.9         & 0.56            & 0.14            \\
RNN-T LID fine-tuned & 10.4 & 5.1 & 3.3 & 0.44 & 0.19 \\
HuBERT          & 9.9 & 4.1 & 3.6 & 1.0 & 0.2 \\
\bottomrule
\end{tabular}

\caption{LID results (EER in \%) for the NIST-LRE-07 tests}
\label{tab:lid}
\end{table}

The results demonstrate that our model performs better for speech with longer durations. Apparently, for longer durations, the ability of the STT encoder to capture phonetic information becomes very useful. However, for shorter durations, the HuBERT model may be able to capture different information, such as an accent. This is likely what helps it identify the language even before enough words are spoken.

\section{Conclusions}
In this paper, we showed that we can extend an RNN-T-based STT model to perform additional tasks, such as emotion classification and language identification. This brings us closer to the goal of a single ASR system that can perform multiple tasks and produce a rich transcript. This system might not be as good in one task as a single-task system, but it still provides significant benefits in term of training, deployment, maintenance, and user experience.

Our results for the emotion classification are currently not as good as other state-of-the-art classifiers. This might be because the original training of the STT model causes it to ignore much of the information in the speech that is useful for correctly identifying the emotions. When we adapt the model for emotion classification, it fails to properly learn how to extract the required information. Additional research is needed to overcome this problem by using a larger training dataset or by combining emotion classification objectives into the original training of the STT.

On the other hand, we found that the RNN-T model is very powerful for language classification and can produce results that match and even surpass previous state-of-the-art models. Additional work is needed to improve the results for short speech segments and for using this ability to create multilingual STT models.


\bibliographystyle{IEEEtran}

\bibliography{mybib}

\end{document}